\documentclass[12pt]{iopart}
\usepackage {graphicx}
\usepackage{iopams}
\usepackage{color}

\begin{document}

\def\be{\begin{equation}}
\def\ee{\end{equation}}

\title[]{Nonlinear dynamics of electron-positron clusters}
\author{Giovanni Manfredi}
\address{Institut de
Physique et Chimie des Mat\'{e}riaux, CNRS and Universit\'{e} de
Strasbourg, BP 43, F-67034 Strasbourg, France}
\ead{giovanni.manfredi@ipcms.unistra.fr}
\author{Paul-Antoine Hervieux}
\address{Institut de
Physique et Chimie des Mat\'{e}riaux, CNRS and Universit\'{e} de
Strasbourg, BP 43, F-67034 Strasbourg, France}
\ead{paul-antoine.hervieux@ipcms.unistra.fr}
\author{Fernando Haas}
\address{Departamento de F\'{\i}sica, Universidade Federal do Paran\'{a}, 81531-990, Curitiba, Paran\'{a}, Brazil}
\ead{ferhaas@fisica.ufpr.br}
\date{\today}

\begin{abstract}
Electron-positron clusters are studied using a quantum hydrodynamic model that includes Coulomb and exchange interactions. A variational Lagrangian method is used to determine their stationary and dynamical properties. The cluster static features are validated against existing Hartree-Fock calculations. In the linear response regime, we investigate both dipole and monopole (breathing) modes. The dipole mode is reminiscent of the surface plasmon mode usually observed in metal clusters. The nonlinear regime is explored by means of numerical simulations. We show that, by exciting the cluster with a chirped laser pulse with slowly varying frequency (autoresonance), it is possible to efficiently separate the electron and positron populations on a timescale of a few tens of femtoseconds.
\end{abstract}

\maketitle

\section{Introduction}\label{sec:intro}
The positron was the first antiparticle to be discovered experimentally (in 1932) and it appears naturally as a negative energy solution of the Dirac equation \cite{Strange}. Positron physics is of great fundamental and practical interest, ranging from condensed matter physics to astrophysics and biological physics.
For instance, positron techniques are useful for the investigation of defects in solids and solid surfaces \cite{Puska}. In medicine, the widespread use of positron-emission tomography (PET) for diagnostics and treatment monitoring requires a sound understanding of the physical and biological effects of positrons on living organisms. In astrophysics and cosmology, understanding the imbalance of matter versus antimatter is one of the major challenges of today's theoretical physics. Finally, recent projects aiming to elucidate the gravitational behavior of antimatter require the careful manipulation of positrons in order to produce anti-hydrogen atoms \cite{gbar}.

Positrons can be  easily obtained from the $\beta^+$ decay of radioactive isotopes, e.g. from $^{22}\rm Na$. The positrons generated in this reaction exhibit a broad energy spectrum that extends up to 540 keV. For practical use in antimatter studies, positrons need to be cooled down to a few electron-volts by means of a moderator and are subsequently stored in a trap \cite{gbar}.
Slow positrons implanted into a porous silica film may efficiently form positronium atoms (Ps) \cite{Liszkay, Cassidy}. Positronium is a bound state constituted of an electron and a positron. It is the lightest particle-antiparticle ``atom", with a relatively long lifetime of 125 ps for the singlet state (para-positronium) and 142 ns for  the triplet state (ortho-positronium).

Positronium may also be viewed as  the simplest ``many-body" electron-positron (e-p) system. For larger numbers of particles, various other states are possible, depending on the density and temperature of the system \cite{yabu}. At high temperatures and low densities, classical e-p plasmas can be formed, both relativistic and nonrelativistic \cite{Thoma, Zank, marklund, shukla}. Lowering the temperature below the Ps ground-state binding energy ($E_{\rm Ps} =6.8~\rm eV$), the electrons and the positrons recombine to form a positronium gas. At still lower temperatures, Ps atoms may even form a Bose-Einstein condensate (BEC) \cite{Platzman}. Because of its light mass, the critical temperature of a Ps gas is much higher than, for instance, that of an alkali atom gas with the same number density, which is obviously an interesting feature from an experimental point of view.
For instance, for a Ps density $n = 10^{24}~ \rm m^{-3}$, the critical temperature would be as high as  $T_c \approx 15 ~K$ \cite{Cassidy2005}.
The realization of a BEC of Ps atoms is a very exciting and challenging project, as such a system could lead to the simultaneous coherent decay of all Ps atoms, thus acting as a powerful gamma ray source.

The critical temperature $T_C \sim \hbar^2n^{2/3}/(m k_B)$ (where $m$ is the electron mass) is attained when the de Broglie thermal wavelength of the Ps atoms, $\lambda_B =\hbar/\sqrt{mk_B T}$, becomes comparable to the mean interparticle distance, measured for instance by the Wigner-Seitz radius $r_s = (4\pi n/3)^{-1/3}$. At even higher densities, when the interparticle distance is of the order of the Ps  ground-state radius $2a_0$ (where $a_0$ is the Bohr radius), the e-p system effectively behaves as a two-component degenerate Fermi liquid \cite{yabu}. This is the ``metallic" phase of electron-positron matter.
For a wide range of values of $r_s$, the ground-state properties of e-p infinite matter were studied by Boronski and Nieminem \cite{boro} using two-component density-functional theory (DFT) within the local density approximation (LDA).

However, it is well known that finite-size metallic systems can also exist. These systems -- known as metal clusters \cite{Ekardt, Brack, deheer} -- are usually composed of $10 \leq N \leq 10^4$ ions and an equal number of electrons (although charged clusters have also been studied). In many respects, they behave as giant ``atoms" where the positive charge is not localized in the nucleus but is distributed more or less uniformly within the cluster.
For metal clusters, the positive charges are ions, whose mass is thousands of times that of the electrons. Thus, it is appropriate, when studying effects occurring on a short timescale ($<100$ fs), to assume that the ions are effectively immobile. Further, for large clusters the ions can be modeled by a uniform positive charge density (jellium approximation).
The electronic ground state, obtained by DFT methods, reveals a shell structure with discrete energy levels, akin to those of ordinary atoms.

Recently, it has been suggested \cite{yatsy, solov} that clusters made of electrons and positrons could also exist. Using either two-component DFT or Hartree-Fock methods, it was shown that such e-p clusters have ground-state densities similar to that of metals ($r_s/a_0 \approx 3.5$) and also display an electronic shell structure. Of course, for e-p clusters it is not possible to use the jellium approximation, as both species have the same mass and are thus equally mobile.
For the same reason, e-p clusters are locally neutral in their ground state, whereas metal clusters display a ``spill-out" effect \cite{Brack}, whereby the electron density extends slightly further than the ion density.

The experimental realization of stable e-p clusters is still ahead of us, mainly because extremely high densities are required, $n \approx 10^{28}~ \rm m^{-3}$ (at such densities, the Fermi temperature is very high, $T_F \approx 10^4 ~\rm K$, so that the e-p gas is degenerate even at room temperature).
However, remarkable advances have been made in the confinement and cooling of positron plasmas. Temperatures smaller than 5 K and densities larger than  $10^{16}~ \rm m^{-3}$ have been achieved in recent years \cite{Jelenkovic, Jorgensen}.
Further, recent technological
progress is opening up the possibility of employing intense laser radiation to
trigger physical processes beyond atomic-physics energy scales, such as electron-positron pair production at high densities \cite{Piazza, Marklund}.
Indeed, very recently, in realistic simulations of a 10 PW laser striking a solid target, it was demonstrated that a maximum positron density of $10^{26}  \rm m^{-3}$ can be obtained, seven orders of magnitude greater than achieved in previous experiments \cite{Ridgers}.

On the other hand, dense gases of interacting Ps atoms have been created by irradiating a thin film of
nanoporous silica with intense positron bursts \cite{Cassidy2005}, reaching densities of the order of $n \approx 10^{21}~ \rm Ps/m^{3}$, just three orders of magnitude lower than the density needed to form a BEC of Ps atoms with $T_C \approx 15 ~K$.
All in all, both Ps BECs and e-p clusters represent admissible states of electron-positron matter under extreme conditions of temperature and density
and in that respect they deserve to be properly investigated theoretically.

As of today, no results exist on the dynamics of e-p clusters, either in the linear or nonlinear regimes. In contrast, the linear response of metal clusters has been the subject of intense investigations in the last few decades \cite{Ekardt85, guet}. A strong dipole resonance is observed near the Mie frequency, which for spherical clusters in the jellium approximation is given by the bulk plasma frequency divided by $\sqrt{3}$. Using more sophisticated approaches, it can be shown that the resonant frequency actually depends on the cluster size. The nonlinear electronic response was investigated more recently by means of phase-space or hydrodynamic methods \cite{Calvayrac}.

In the present work, we aim at characterizing the linear and nonlinear response of e-p clusters. We use a variational approach based on a two-component quantum hydrodynamic method that incorporates the kinetic energy, the Coulomb interaction, and the exchange energy, but neglects higher-order correlations (Sec. \ref{sec:model}). Thus, our method can be viewed as an approximation of the two-component Hartree-Fock equations. Only one adjustable parameter (related to the gradient correction of the exchange energy) appears in our model and is determined by matching our solution for the ground state (Sec. \ref{sec:steady}) with that obtained through Hartree-Fock calculations. Subsequently, we study the linear response of the e-p clusters, which reveals both dipole and monopole resonances (Sec. \ref{sec:linear}). Finally, we investigate numerically the nonlinear dynamics and show that the electron and positron populations can be effectively separated on a timescale much shorter than that of mutual annihilation (Sec. \ref{sec:autores}).
Conclusions are presented in Sec. \ref{sec:conclusion}.

\section{Quantum hydrodynamic model and Lagrangian method}\label{sec:model}
In our approach, the electron-positron system is governed by a set of quantum hydrodynamic equations for the densities $n_i$ and the mean velocities ${\bf u}_i$, where the subscript $i = e,p$ denotes each species \cite{manfredi_hydro}. Hydrodynamic methods have been used successfully in the past to model the electron dynamics in molecular systems \cite{Brew}, metal clusters and nanoparticles \cite{Bane, Domps}, thin metal films \cite{Crouseilles}, and quantum plasmas \cite{eliasson, Shukla}.
In the following, we will always use atomic units (a.u.)
such that space is normalized to the Bohr radius $a_0=4\pi\varepsilon_0\hbar^2/(me^2)$, energy to the Hartree energy $E_H=me^4/(4\pi\varepsilon_0\hbar)^2$, and time to $\tau_H=\hbar/E_H$.
In a.u., the hydrodynamic equations read as follows:
\begin{eqnarray}
\label{continuity}
\frac{\partial n_i}{\partial t} &+& \nabla\cdot\,(n_i{\bf u}_i) = 0 \,,\\
 \frac{\partial\,{\bf u}_i}{\partial\,t} &+& {\bf u}_{i}\cdot\,\nabla\,{\bf u}_i  = - \frac{\nabla p_i}{n_i} - q_i \nabla V_H \nonumber
 - \nabla V_{X,i}
 \label{momentum}
 + \frac{1}{2} ~\nabla
\left(\frac{\nabla^2 \sqrt{n_i}}{\sqrt{n_i}} \right) ,
\end{eqnarray}
where $q_i = \pm 1$ for positrons and electrons.
The four terms on the right-hand side of Eq. (\ref{momentum}) represent respectively the pressure, the Hartree (Coulomb) potential, the exchange potential, and the von Weizs\"{a}cker correction (sometimes referred to as the Bohm potential).

We neglect correlations altogether in our model. This is done for two main reasons: first, the correlation functional for finite-size systems of same-mass particles is extremely complicated, particularly the cross terms (electron correlations due to the positron cloud, and vice-versa) \cite{boro}; second, results that only include the exchange term can be directly compared to exact calculations performed with the Hartree-Fock equations \cite{yatsy}.

The Hartree potential $V_H$ satisfies Poisson's equation
\begin{equation}
\nabla^2 V_H = 4\pi(n_e - n_p) \,.
\label{eq:poisson}
\end{equation}
The exchange potential is derived from the exchange energy functional
\be
E_X[n_i] =  -\frac{3 (3\pi^2)^{1/3} }{4\pi} \int  n_{i}^{4/3} d\mathbf{r}\, -
\beta \int \frac{(\nabla n_i)^2}{n_{i}^{4/3}} d\mathbf{r} ,
\label{ex_energy}
\ee
where the first term is the usual LDA expression and the second term is a gradient correction \cite{becke}. The latter is on the same level of approximation as the von Weizs\"{a}cker correction to the kinetic energy. The parameter $\beta$ will be determined numerically by comparing our results to exact Hartree-Fock calculations (see Sec. \ref{sec:steady}). The obtained value $\beta = 0.0135$ is larger than the best-fit value usually employed in atomic-structure calculations, which is $\beta \approx 0.005$ \cite{becke}.
The exchange potential is then obtained as the functional derivative of the exchange energy with respect to the density: $V_{X, i} = \delta E_X[n_i]/\delta n_i$.

Finally, for the pressure we use the expression of the Fermi pressure for a zero-temperature electron (or positron) gas:
\begin{equation}
p_i = \frac{2}{5}n_i E_{F}[n_i] = \frac{1}{5} (3\pi^2)^{2/3}n_{i}^{5/3} \,,
\label{pressure}
\end{equation}
where $ E_{F}[n_i]$ is the Fermi energy. Since the ground state density of e-p clusters is similar to that of metals (see next section), their Fermi temperature is of the order $T_F \sim 10^4 ~\rm K$, which justifies the zero-temperature assumption.

It can be shown that the hydrodynamic equations (\ref{continuity})-(\ref{momentum}) can be derived from the following Lagrangian density ${\cal L}$ \cite{haas_breath}:
\begin{eqnarray}
{\cal L} &=& \sum_{i=e,p} \Big\{ \frac{n_i }{2}\left(\nabla S_i\right)^2 + n_i \frac{\partial S_i}{\partial t} +  \frac{\left(\nabla n_i\right)^2}{8n_i} + \frac{3}{10}(3\pi^2)^{2/3} n_{i}^{5/3} \nonumber  \\
 &-&  \frac{3 (3\pi^2)^{1/3} }{4\pi} n_{i}^{4/3} - \beta \frac{(\nabla n_i)^2}{n_i^{4/3}} \Big\}  -
\frac{(\nabla V_H)^2}{8\pi} + (n_p - n_e)V_H ~,
\label{lag_density}
\end{eqnarray}
where the independent fields are $n_i$, $S_i$, and
$V_H$. The velocity fields ${\bf u}_i$ follow from the auxiliary functions
$S_i = S_{i}({\bf r},t)$ through ${\bf u}_i = \nabla S_i$.

Our purpose now is to derive -- using the variational approach detailed in Ref. \cite{haas_breath} -- a set of evolution equations for a small number of macroscopic quantities that characterize the electron and positron density profiles. With this aim in mind, we assume that the density profiles are Gaussian functions
\begin{equation}
\label{gauss}
n_i ({\mathbf r},t) = \frac{N}{\pi^{3/2} \sigma_{i}^3}\,\exp\left[- \frac{x^2+y^2+(z-d_i)^2}{\sigma_{i}^2}\right] \,,
\end{equation}
where $N$ is the total number of electrons and positrons (assuming overall charge neutrality), whereas  $\sigma_i(t)$ and $d_i(t)$ are time-dependent functions defining respectively the size of the electron and positron clouds and the displacement in the $z$ direction. We allow for a displacement along the $z$ axis because we will later suggest using a laser pulse to excite the dipole mode (see Secs. \ref{sec:linear}-\ref{sec:autores}).

Of course, such a Gaussian {\it Ansatz} is not exact and may even differ significantly from the electron density obtained, for instance, from a HF calculation. Nevertheless, it is a useful and relatively safe procedure to obtain a mathematically treatable set of equations that can be solved either exactly or with minimal numerical effort. For instance, it was noticed in Ref. \cite{haas_breath} that, even when the actual density is not well approximated by a Gaussian profile, the resonant frequencies computed with our technique are still very close to the exact ones.

For the above density profiles, the exact solution of Poisson's equation (\ref{eq:poisson}) is
\begin{equation}
V_H =\sum_{i=e,p} \frac{N q_i}{s} \,{\rm Erf}\left[\frac{s}{\sigma_i}\right] \,,
\end{equation}
where ${\rm Erf}(s)$ is the error function and $s^2=x^2+y^2+(z-d_i)^2$.
In addition, the continuity equation (\ref{continuity}) is exactly solved by the following velocity field
\begin{equation}
\label{u}
{\bf u}_i = \frac{\dot\sigma_i}{\sigma_i}\,(x\hat{x} + y\hat{y}) + \left[\frac{\dot\sigma_i}{\sigma_i}\,(z-d_i) + \dot{d}_i\right]\hat{z} \,,
\end{equation}
with
\begin{equation}
\label{phase}
S_i = \left[\frac{\dot\sigma_i}{2\sigma_i} (x^2+y^2+[z-d_i]^2) + \dot{d}_{i}(z-d_i)\right],
\end{equation}
where the dot denotes derivation with respect to time and $\hat{x}$, $\hat{y}$, $\hat{z}$ are unit vectors along each direction. An irrelevant additive function of time was ignored in Eq. (\ref{phase}).

We can now compute the Lagrangian $L = -N^{-1} \int{\cal L}\,d{\bf r}$,
where the multiplicative factor was introduced for convenience of notation.
In atomic units, the Lagrangian reads as:
\begin{eqnarray}
L &=& \frac{3}{2} \sum_{i=e,p}\Big\{\frac{\dot{d}_{i}^2}{3}
+ \frac{\dot{\sigma}_{i}^2}{2} - \frac{1}{2\sigma_{i}^2} - \frac{C_K N^{2/3}}{\sigma_i^2}- \frac{C_H N}{\sigma_i} \nonumber \\
&+&  \frac{C_X N^{1/3}}{\sigma_i} + \frac{C_X'}{N^{1/3}\sigma_i}
 \Big\} +
N\,{\rm Erf}
\left(\frac{d_e - d_p}{\sqrt{\sigma_{e}^2+\sigma_{p}^2}}\right) \,,
\label{lag_au}
\end{eqnarray}
where the coefficients
\begin{eqnarray}
C_K &=& \left(\frac{3}{5}\right)^{3/2} \frac{(3\pi^2)^{2/3}}{5\pi}\approx  0.2832, \nonumber\\
C_H &=& \frac{\sqrt{2}}{3\sqrt{\pi}}  \approx 0.2660, \nonumber \\
C_X &=& \frac{3}{16} \frac{3^{5/6}}{\pi^{5/6}} \approx 0.1804, \nonumber\\
C_X' &=&  9\beta\sqrt{3\pi/2} \approx 0.2638 ~~~ \rm (for ~ \beta = 0.0135), \nonumber
\end{eqnarray}
correspond respectively to the kinetic energy (Fermi pressure), the Hartree energy, and the exchange energy.

The corresponding equations of motion are obtained from the standard Euler-Lagrange equations
\be
\frac{d}{d t} \frac{\partial L}{\partial {\dot \zeta}} -
 \frac{\partial L}{\partial \zeta} =0~,
\ee
where $\zeta = \{d_e, d_p, \sigma_e, \sigma_p \}$. The result is:
\begin{eqnarray}
\ddot{d}_i &=& \frac{N}{|d_e-d_p|^2}\,\Big[\frac{2}{\pi^{1/2}}
\frac{d_i-d_j}{\Sigma}\exp\left(- \frac{(d_e - d_p)^2}{\Sigma^2}\right)
- {\rm Erf}\left(\frac{d_e - d_p}{\Sigma}\right) \Big]  ,\\
\ddot\sigma_i &=&
\frac{1}{\sigma_{i}^3} +
\frac{2 C_K N^{2/3}}{\sigma_i^3} +
\frac{C_H N}{\sigma_i^2} -
\frac{C_X N^{1/3}}{\sigma_i^2} \nonumber \\
&-& \frac{C_X'}{\sigma_i^2 N^{1/3}} - \frac{4N}{3 \pi^{1/2}} \exp\left(- \frac{(d_e - d_p)^2}{\Sigma^2}\right) \frac{\sigma_i}{\Sigma^3} \,,
\end{eqnarray}
where we have defined $\Sigma^2 = \sigma_{e}^2+\sigma_{p}^2$.

It is preferable to use center-of-mass and relative coordinates, defined as $D=(d_e+d_p)/2$ and $d=d_e-d_p$ . We obtain that $\ddot D=0$ and
\begin{eqnarray}
\ddot{d} &=& \frac{2N}{d^2}\left\{\frac{2d}{\sqrt{\pi}\Sigma}\exp\left(- \frac{d^2}{\Sigma^2}\right) - \,{\rm Erf}\left(\frac{d}{\Sigma}\right)\right\} \label{eq_dipole}\\
\ddot\sigma_i &=&
\frac{1}{\sigma_{i}^3} +
\frac{2 C_K N^{2/3}}{\sigma_i^3} +
\frac{C_H N}{\sigma_i^2} - \frac{C_X N^{1/3}}{\sigma_i^2} \nonumber \\ &-& \frac{C_X'}{\sigma_i^2  N^{1/3}}
- \frac{4N}{3 \sqrt{\pi}} \exp\left(- \frac{d^2}{\Sigma^2}\right)
\frac{\sigma_i}{\Sigma^3} \,.
\label{eq_sigma}
\end{eqnarray}
In the next sections, the above equations will be used to study the steady state properties, as well as the linear and nonlinear responses of e-p clusters.

\section{Steady states}\label{sec:steady}
Yatsyshin et al. \cite{yatsy} and Solovyov et al. \cite{solov} have investigated the stationary properties of electron-positron clusters using respectively Hartree-Fock (HF) and DFT calculations. In all cases, they obtain neutral states where the electron and positron density profiles are locally identical.

Within the framework of our model, we also look for neutral equilibria for which $\sigma_e=\sigma_p=\sigma_0$ and $d=0$. Substituting into Eq. (\ref{eq_sigma}), the Hartree terms cancel out, because the equilibrium is locally neutral. We obtain:
\begin{equation}
0 =  \frac{1}{\sigma_{0}^3} +
\frac{2 C_K N^{2/3}}{\sigma_0^3} -
\frac{C_X N^{1/3}}{\sigma_0^2}- \frac{C_X' N^{-1/3}}{\sigma_0^2}.
\end{equation}
The steady state is then given by a Gaussian density with a width equal to
\begin{equation}
\sigma_0 = \frac{1+ 2C_K \, N^{2/3}}{C_X N^{1/3}+C_X' N^{-1/3}}.
\label{sigma_0}
\end{equation}
The corresponding Wigner-Seitz radius $r_s$ can be defined using the peak value of the density:
\be
\frac{4}{3}\pi r_s^3 = n_{\rm peak}^{-1} \equiv \frac{\pi^{3/2}\sigma_0^3}{N}.
\ee

So far, our model still contains a free parameter, namely the coefficient $\beta$ appearing  in the gradient correction of the exchange energy functional, Eq. (\ref{ex_energy}). For atomic systems, this parameter has usually been determined by comparison with exact HF calculations, yielding a value $\beta \approx 0.005$ \cite{becke}. Here, we follow the same procedure and compare our analytical result, Eq. (\ref{sigma_0}), with the HF calculations published in Ref. \cite{yatsy}. It turns out that the best fit on the Wigner-Seitz radius is obtained for $\beta = 0.0135$ and we will retain this value for all forthcoming calculations. Of course, oscillations of $r_s$ with the system size $N$ -- which are  due to shell effects and are present in the HF calculations -- cannot be recovered with our simple model.

\begin{figure}[ht]
\centering
\includegraphics[width=0.9\textwidth]{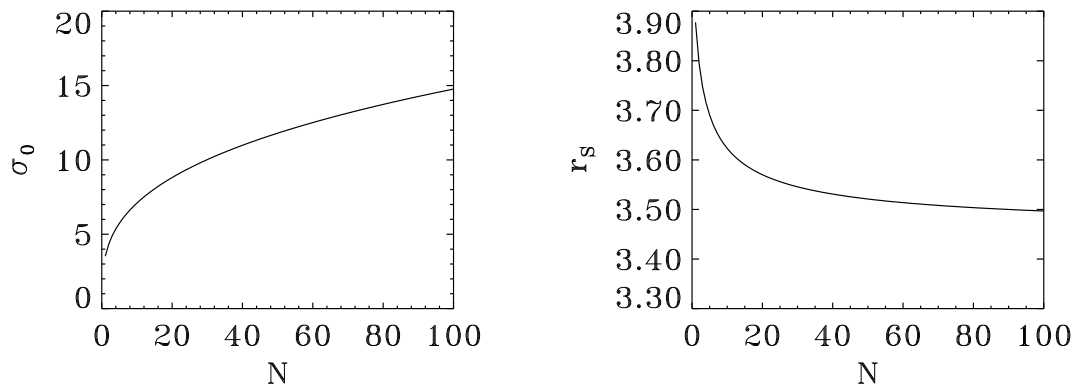}
\caption{Width of the e-p cluster (left) and corresponding Wigner-Seitz radius (right) in atomic units, as a function of the number of electrons/positrons $N$.} \label{fig:steady}
\end{figure}

The system size $\sigma_0$ and the Wigner-Seitz radius are shown in Fig. \ref{fig:steady} as a function of the number of particles.  As expected, $\sigma_0$ goes like $N^{1/3}$ for large $N$, whereas $r_s/ a_0$ varies between 3.45 in the bulk ($N \gg 1$) and 3.9 for small values of $N$ (although it should be pointed out that the validity of our approach becomes questionable for very small systems). In addition, the bulk value of $r_s$ does not depend on the chosen value of $\beta$.

These findings are in accordance with standard results obtained for metal clusters \cite{Calvayrac}. This is important, as one can envision to excite such e-p systems with the same optical means that are currently used to study the dynamical response of metal clusters.

\section{Linear response}\label{sec:linear}
We now investigate the linear response of the e-p cluster under external excitations. Two types of mode can be studied in the framework of our model: (i) a dipole mode, where the electron and positron clouds oscillate with respect to each other and (ii) a breathing or monopole mode \cite{haas_breath, bauch}, where the radii $\sigma_i$ of both clouds oscillate, either in phase or in antiphase.
It is important to stress that, at the level of the linear response, the dipole and breathing modes are completely decoupled, although of course some coupling will occur in the nonlinear regime.

\subsection{Dipole mode}\label{subsec:dipole}
\begin{figure}[ht]
\centering
\includegraphics[width=0.4\textwidth]{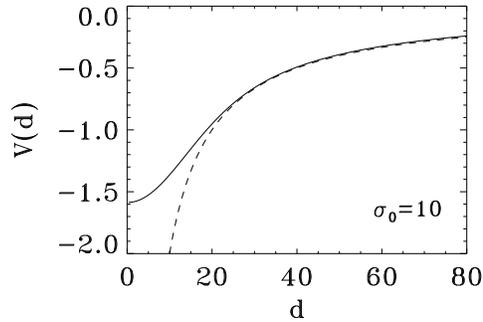}
\caption{Effective potential (solid line) and standard Coulomb potential in vacuum (dashed line), for a case with $\sigma_0=10$.} \label{fig:poten}
\end{figure}

Let us rewrite the equation for the dipole $d(t)$, assuming that $\sigma_e=\sigma_p=\sigma_0$:
\be
\ddot{d} = \frac{2N}{d^2}\left\{\sqrt{\frac{2}{\pi}} \frac{d}{\sigma_0}\exp\left(- \frac{d^2}{2\sigma_0^2}\right) - \,{\rm Erf}\left(\frac{d}{\sqrt{2}\sigma_0}\right)\right\}.
\label{eq_dipole_0}
\ee
The right-hand side of Eq. (\ref{eq_dipole_0}) can be written as $-\partial V/\partial d$, which implicitly defines the effective potential $V(d)$. On Fig. \ref{fig:poten} we plot the effective potential together with the standard Coulomb potential in vacuum: $V_{\rm Coul}(d) = -2N/d$. The factor $2N$ is the total number of charges (electrons and positrons).
We can see that the effective potential is a ``regularized'' version of the Coulomb potential: the divergence at $d=0$ has disappeared and we are left with a smooth potential well.

We now linearize Eq. (\ref{eq_dipole_0}) around the equilibrium $d=0$, using the expansion ${\rm Erf}(x) = (2/\sqrt{\pi})(x-x^3/3 +\dots)$. We obtain $\ddot d+ \Omega_d^2 d=0$, with the linear frequency:
\be
\Omega_d^2= \frac{2}{3}\, \sqrt{\frac{2}{\pi}}\, \frac{N}{\sigma_0^3},
\ee
or in SI units:
\be
\Omega_d^2= \frac{2}{3}\, \sqrt{\frac{2}{\pi}}\, \frac{N}{\sigma_0^3}
\frac{e^2}{4\pi m\varepsilon_0},
\ee

We now define the average density as \cite{haas_breath}
\be
\langle n\rangle  \equiv \frac{\int n^{2} ({\mathbf r}) d{\mathbf r}}{\int n({\mathbf r}) d{\mathbf r}}= \frac{N}{(2\pi)^{3/2}\sigma_0^3}
\ee
and the plasma frequency computed with the reduced mass $\bar{m}=m/2$:
\be
{\bar \omega}_p^2 = \frac{e^2 \langle n\rangle}{\bar{m} \varepsilon_0} = \frac{2e^2 N}{(2\pi)^{3/2}\sigma_0^3 m \varepsilon_0}.
\ee
The linear dipole frequency then becomes:
\be
\Omega_d = \frac{{\bar \omega}_p}{\sqrt{3}}~,
\ee
which is the same as the Mie frequency for spherical metal clusters \cite{Brack}.

\subsection{Breathing modes}\label{subsec:breath}
We linearize Eqs. (\ref{eq_dipole})-(\ref{eq_sigma}) around the equilibrium $\sigma_e=\sigma_p=\sigma_0$ and $d=0$, by writing $\sigma_i=\sigma_0 + {\tilde \sigma_i}$ and $d = {\tilde d}$, with
${\tilde \sigma_i}, {\tilde d} \ll \sigma_0$. We obtain that the equations for ${\tilde d}$ and ${\tilde \sigma_i}$ completely decouple. The former yields the dipole mode described in the preceding subsection, while the equation for ${\tilde \sigma_i}$ reads as:
\begin{eqnarray}
\ddot{\tilde {\sigma_i}} &=&
 -\frac{3 {\tilde \sigma_i}}{\sigma_0^4} - 6C_K N^{2/3}\frac{{\tilde \sigma_i}}{\sigma_0^4} + \frac{N}{\sqrt{2\pi}} \frac{{\tilde \sigma_e} + {\tilde \sigma_p}}{\sigma_0^3} \nonumber \\
&-& (3C_H N-2C_X N^{1/3}-2 C_X' N^{-1/3}) \frac{{\tilde \sigma_i}}{\sigma_0^3}
 .
\label{eq_sigma_lin}
\end{eqnarray}
By Fourier transforming in the time variable, i.e., replacing $\ddot{\tilde {\sigma_i}}$ with $-\Omega^2 {\tilde \sigma_i}$, the linearized equations can be written as follows:
\begin{eqnarray}
A {\tilde \sigma_e} &+& B {\tilde \sigma_p} = 0, \nonumber \\
B {\tilde \sigma_e} &+& A {\tilde \sigma_p} = 0, \nonumber
\label{eq_sigma_lin2}
\end{eqnarray}
where
\begin{eqnarray}
A(\Omega) &=& \Omega^2-\frac{6}{\sigma_0^4}({1\over 2}+C_K N^{2/3}) -\frac{N}{\sqrt{2\pi}\sigma_0^3}  \nonumber \\
&+& \frac{2C_X N^{1/3}}{\sigma_0^3}
+ \frac{2C_X' N^{-1/3}}{\sigma_0^3} \\
B &=& \frac{N}{\sqrt{2\pi}\,\sigma_0^3}.
\end{eqnarray}

The relevant dispersion relation can be written as $A(\Omega)=\pm B$, which yields the two resonant frequencies:
\begin{eqnarray}
\Omega_{-}^2 &=& \frac{6}{\sigma_0^4}({1\over 2}+C_K N^{2/3})-\frac{2C_XN^{1/3}}{\sigma_0^3} - \frac{2C_X'}{\sigma_0^3 N^{1/3}}\\
\Omega_{+}^2 &=& \Omega_{-}^2+ \sqrt{\frac{2}{\pi}} \frac{N}{\sigma_0^3} \,.
\end{eqnarray}
In the large $N$ limit, $\sigma_0 \to (2C_K/C_X)\, N^{1/3}$, and we obtain
\begin{eqnarray}
\Omega_{-}^2 &=& C_X \left(\frac{C_X}{2C_K}\right)^3 N^{-2/3} \to 0\\
\Omega_{+}^2 &= & \sqrt{\frac{2}{\pi}} \frac{N}{\sigma_0^3} = 4\pi \langle n \rangle = \omega_p^2 = \frac{{\bar \omega}_p^2}{2}\,.
\end{eqnarray}

We note that the solution $\Omega_{-}$ corresponds to $A=-B$ and therefore ${\tilde \sigma_e} = {\tilde \sigma_p}$: this is a neutral mode where the electron and positron densities fluctuate in phase. In contrast, the solution $\Omega_{+}$ corresponds to  ${\tilde \sigma_e} = -{\tilde \sigma_p}$: it is a nonneutral mode, with the density fluctuations oscillating in antiphase.
We further note that for the neutral mode $\Omega_{-}$ only the Bohm, exchange and Fermi pressure terms play a role: these are all {\em local} terms, so that the mode disappears for very large clusters. In contrast, the nonneutral mode $\Omega_{+}$ depends on the Hartree potential, which is nonlocal, hence this mode persists for all cluster sizes.

\section{Nonlinear response and autoresonant excitation}\label{sec:autores}
We now turn our attention to the excitation of the electron and positron dynamics by means of electromagnetic waves (laser pulses). First, it should be noted that the relevant linear frequencies computed in the preceding section are of the order of a few electron-volts. For instance, for $N=100$, one finds $\Omega_d = 3.50$ eV, $\Omega_+= 4.31$ eV, and $\Omega_{-} = 0.45$ eV. These frequencies fall within the visible or near ultraviolet (UV) spectrum, which is good news, as visible and near-UV lasers are commonly employed in ultrafast optics experiments. For such lasers, the wavelength is several hundred nanometers long, i.e., much larger than the size of a typical e-p cluster (see Fig. \ref{fig:steady}). This means that only the dipole mode can be excited directly, just like for ordinary metal clusters. The only hope to excite the breathing modes is via nonlinear coupling to the dipole mode.

In order to model the interaction of an electromagnetic wave with our e-p system, we introduce an external homogeneous electric field parallel to the $z$ direction, ${\bf E}=E(t) \hat{z}$. This amounts to adding a term $-2E(t)$ on the right-hand side of Eq. (\ref{eq_dipole}) (the prefactor $-2$ appears because the force is $-E$ for the electrons and $+E$ the for positrons).
Since the system is globally neutral, a homogeneous field does not affect the center-of-mass equation of motion, which remains $\ddot D = 0$.
Notice that this procedure is completely consistent with our Lagrangian approach and could have been obtained rigourously from the start by adding an external energy term $-\sum_{i}q_i z E(t) n_i$ to the Lagrangian density in Eq. (\ref{lag_density}).

The idea here is to excite the dipole mode in order to separate the electron and positron populations using an oscillating dipolar electric field. Of course, this can always be achieved by using a sufficiently strong field, comparable to the electric field that binds the electrons and positrons together, which is of the order of 1 a.u. $= 5.14 \times 10^{11}$ V/m.

One could hope to lower the required field by exciting the system at the resonant dipole frequency, i.e. $E(t) =  E_0 \cos(\omega_0 t)$ with $\omega_0=\Omega_d$. However, the effective potential between the electron and positron clouds is not harmonic, as is apparent from Fig. \ref{fig:poten}. As the distance $d(t)$ grows under the influence of the resonant field, the system increases its energy and eventually reaches the anharmonic region of the confining potential. At this point, the laser frequency will no longer match the energy-dependent frequency of the confining potential, so that
the resonance condition is lost and absorption of the laser light
becomes inefficient.

\begin{figure}[ht]
\centering
\includegraphics[width=0.8\textwidth]{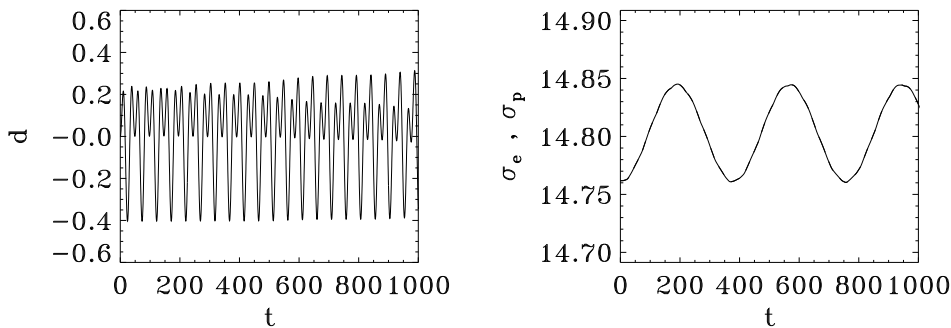}
\caption{Evolution of the dipole $d$ (left frame) and widths $\sigma_i$ (right frame) for an e-p cluster with $N=100$. The excitation frequency $\omega_0$ is constant and equal to $\Omega_d$ and the amplitude is $E_0= 0.005$ a.u.} \label{fig:nonauto}
\end{figure}

This is apparent from Fig. \ref{fig:nonauto}, where we show a numerical solution of the fully nonlinear equations (\ref{eq_dipole})-(\ref{eq_sigma}) obtained with a second-order leap-frog method, for a cluster with $N=100$. The distance $d$ between the electron and positron clouds always remains much smaller than the size $\sigma_i$ of each cloud, so that no separation is achieved. We also note that the neutral breathing frequency $\Omega_{-}$ is excited nonlinearly; indeed, both $\sigma_e$ and $\sigma_p$ oscillate in phase (they are indistinguishable on the figure) with a period very close to $2\pi/\Omega_{-} = 377$ a.u.

The above limitation can be overcome by
resorting to {\em autoresonant} excitation \cite{Fajans}.
Basically, autoresonance occurs when a classical
nonlinear oscillator is externally excited by an oscillating field
with slowly varying frequency. In our notation
\begin{equation}
E(t) = E_0 g(t) \cos\left[\omega_0(t-t_0)+\frac{1}{2} \alpha
(t-t_0)^2\right], \label{autoforce}
\end{equation}
where $E_0$ is the excitation amplitude, $g(t)$ is a Gaussian envelope function with peak value equal to unity, and $\omega_0$ is equal to $\Omega_d$ in our case. For instance, when $\alpha<0$, the time-dependent
frequency $\omega(t) = \omega_0+\alpha (t-t_0)$ is initially
larger than the linear frequency, reaches $ \omega_0$ at
$t=t_0$, and then goes on slowly decreasing  with a rate equal to
$\alpha$. It can be shown that, for $|\alpha| \ll \omega_0^2$ and
$E_0$ above a certain threshold, the instantaneous oscillator
frequency becomes ``locked" to the instantaneous excitation
frequency, so that the resonance condition is always satisfied. In
that case, the amplitude of the oscillations grows indefinitely
and without saturation, until of course some other effect kicks
in. It was previously shown \cite{Fajans} that the threshold
behaves as $E_0^{\rm th} \sim \omega_0^{1/2} |\alpha|^{3/4}$, implying that the
amplitude can be arbitrarily small, provided that the
external frequency varies slowly enough. Autoresonant
excitation has been been fruitfully applied to several systems,
including charged antiparticles \cite{andresen}, the quantum pendulum \cite{barth, murch}, and semiconductor quantum dots \cite{manfredi_auto}.

We now apply an autoresonant excitation to an e-p cluster with $N=100$, using the same amplitude as in Fig. \ref{fig:nonauto}, $E_0 = 0.005$ a.u., and $\alpha = -10^{-4}$ ($\alpha$ must be negative because here the frequency is a decreasing function of the energy).
The envelope function $g(t)=\exp[-(t-t_0)^2/2\Delta^2]$ reaches its maximum at $t_0$, i.e., the instant when the external excitation frequency coincides with the linear resonant frequency $\Omega_d$. The width of the pulse is $\Delta=740$ a.u. $\approx 18$ fs, which is a realistic duration for current femtosecond laser pulses.

\begin{figure}[ht]
\centering
\includegraphics[width=0.8\textwidth]{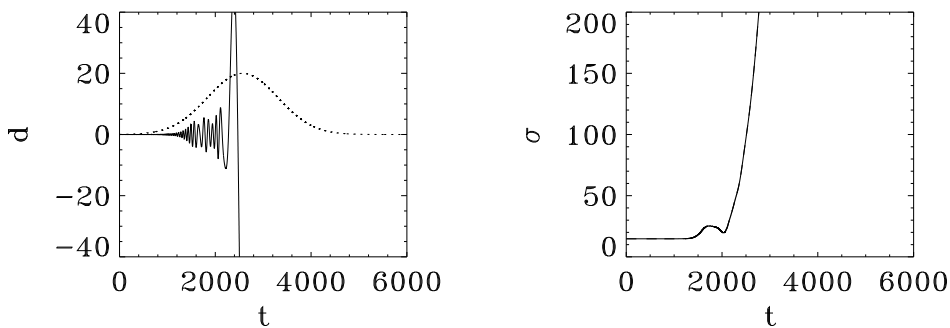}
\caption{Evolution of the dipole $d$ (left frame) and widths $\sigma_i$ (right frame) for an e-p cluster with $N=100$. The excitation is autoresonant with $E_0= 0.005$ and $\alpha = -10^{-4}$ a.u. The dotted line on the left frame represents, in arbitrary units, the electric field envelope $g(t)$.} \label{fig:auto_dip}
\end{figure}

The results of fully nonlinear numerical calculations for the autoresonant case are shown in Fig. \ref{fig:auto_dip}. With the same field amplitude as before, we observe that both the dipole $d(t)$ (distance between the centers of the two clouds) and the widths $\sigma_i$ grow to very large values. This is a clear sign that the autoresonant technique allows us to go well beyond the resonant excitation at the linear frequency.

Nevertheless, since both $d$ and $\sigma_i$ keep growing,
the {\em overlap} between the electron and positron densities is not necessarily decreasing with time. This overlap is the truly interesting quantity, because it tells us whether the two populations are well separated or not. The overlap is also proportional to the probability of e-p annihilation, which we have neglected so far but may play a role over longer times.

\begin{figure}[ht]
\centering
\includegraphics[width=0.8\textwidth]{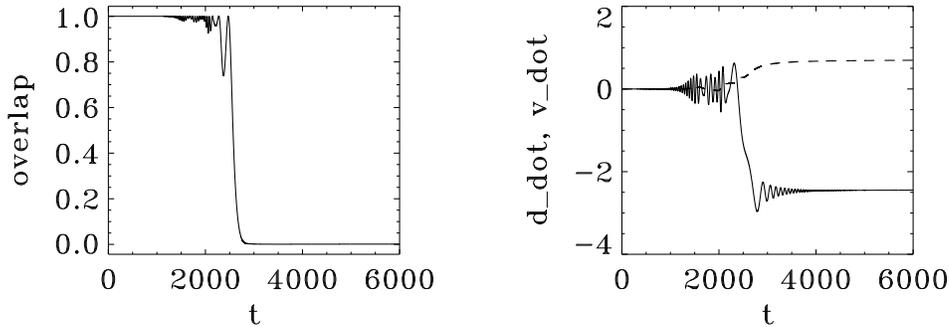}
\caption{Evolution of the overlap $I(t)$ (left frame) and the velocities $\dot d$ (bottom frame, solid line) and $\dot \sigma_i$ (right frame, dashed line) for an e-p cluster with $N=100$. The excitation is autoresonant with $E_0= 0.005$ and $\alpha = -10^{-4}$ a.u.} \label{fig:auto_over}
\end{figure}

We define the normalized overlap $I(t)$ as:
\be
I(t) = \frac{\int n_e n_p~ d{\mathbf r}} {\int n_0^2 ~d{\mathbf r}}~,
\label{overlap}
\ee
where $n_0({\mathbf r})$ is the initial equilibrium density, so that $I(0)=1$. If we assume that $\sigma_e(t) = \sigma_p(t)$ (which was observed to be true for all cases that we simulated), then the overlap can be computed analytically and yields: $I(t)=\exp(-d^2/2 \sigma_{e,p}^2)$. In Fig. \ref{fig:auto_over} we plot the time history of the overlap and of the ``velocities" $\dot d$ and $\dot \sigma_i$. The overlap quickly drops to zero as soon as the autoresonant mechanism becomes effective, which confirms that the two species do become well separated.
The evolution of  $\dot d$ and $\dot \sigma_i$ shows that both the dipole and the widths grow linearly in time but with different velocities, with $|\dot d| > |\dot \sigma_i|$.

\section{Conclusion}\label{sec:conclusion}
In this paper we studied the static and dynamical properties of electron-positron clusters. Our model takes into account quantum and finite-size effects and incorporates the Coulomb forces and exchange interactions. Using a Lagrangian approach and a Gaussian ansatz for the density profiles, we were able to derive a set of ordinary differential equations for the radii of the electron and positron clouds, $\sigma_e$ and $\sigma_p$, and the distance $d$ between their centers of mass.
The only free parameter of the model, $\beta$, related to the gradient correction to the exchange interaction, was determined by matching our static results with those issued from exact Hartree-Fock calculations.

The above approach allowed us to investigate for the first time the dynamical properties of e-p clusters. We first concentrated on the linear response, which revealed three resonant frequencies: a dipole mode (oscillations of $d$) and two breathing modes (oscillations of the $\sigma_i$). For typical parameters, these resonant frequencies lie within or near the visible spectrum.

The dipole mode can be excited with an external oscillating electric field, such as that provided by a laser pulse. However, the resonant excitation rapidly becomes inefficient when the dipole $d$ grows beyond the harmonic part of the confining potential and starts exploring the nonlinear region, where the frequency is energy-dependent.

This drawback was overcome by resorting to autoresonance, whereby the excitation frequency varies slowly during the pulse.
The autoresonant technique allowed us to efficiently separate the electron and positron populations using a laser pulse in the visible -- or near visible -- range, with a peak electric field $E_0 = 0.005$ a.u. = $2.57 \times 10^9$ V/m and pulse duration $\approx 20$ fs. These values are largely independent on the cluster size $N$ and may be achieved
experimentally using current ultrafast spectroscopy techniques.

Finally, it is important to stress that the species separation could be achieved in very short times ($\approx 20$ fs). This is much shorter than the lifetime of electrons and positrons in a positronium atom, which is 125 ps for the singlet state (para-positronium) and 142 ns for  the triplet state (ortho-positronium).
On the other hand, for a nonrelativistic e-p plasma (free e-p pairs), the annihilation rate $\gamma_D$ can be estimated with the Dirac formula \cite{Dirac, Gribakin}: $\gamma_D = \pi r_0^2 c n$, where $r_0$ is the classical electron radius and $c$ is the speed of light in vacuum.
Using the typical density of an e-p cluster, this yields $\gamma_D \approx 3\times 10^{8} \rm s^{-1}$, i.e., an annihilation event roughly every 3 ns.
This is again much longer than the separation time computed above.

Of course, the e-p lifetime in bound clusters like those studied in this work may differ significantly from the values observed for positronium atoms or for free e-p plasmas.
Nevertheless, since the difference between the separation time and the estimated annihilation time is so large (4--5 orders of magnitudes), one may reasonably expect that effective separation can be achieved before annihilation starts playing a significant role.

\section*{Acknowledgements}
This work was partially funded by the Agence Nationale de la Recherche (contract ANR-10-BLAN-0420) and by the Conselho Nacional de Desenvolvimento Cientifico e Tecnologico (CNPq).

\end{document}